\documentclass{article}

\usepackage{arxiv}

\usepackage[utf8]{inputenc} 
\usepackage[T1]{fontenc}    
\usepackage{hyperref}       
\usepackage{url}            
\usepackage{booktabs}       
\usepackage{amsfonts}       
\usepackage{nicefrac}       
\usepackage{microtype}      
\usepackage{lipsum}		
\usepackage{graphicx}
\usepackage{gensymb}
\usepackage{doi}
\usepackage[numbers]{natbib}

\title{Phase Biasing System for Optical Gyroscope Using Passive Non-Reciprocal Polarization Techniques}

\author{
  Onder Akcaalan\\
  Independent Researcher \\
  Hamburg, Germany\\
  \texttt{\ onderakcaalan@gmail.com*} \\
     \And
 Melike Gumus Akcaalan\\
  Institute for Nanostructure and Solid-State Physics\\
  University of Hamburg \\
  Hamburg, Germany\\
  \texttt{\ melike.gumus.akcaalan@uni-hamburg.de} \\
}

\renewcommand{\shorttitle}{\textit{arXiv} Template}

\hypersetup{
pdftitle={A template for the arxiv style},
pdfsubject={q-bio.NC, q-bio.QM},
pdfauthor={David S.~Hippocampus, Elias D.~Striatum},
pdfkeywords={First keyword, Second keyword, More},
}

\begin{document}
\maketitle

\begin{abstract}
Interferometric Fiber Optic Gyroscopes (IFOGs) are widely used in precision navigation systems due to their high sensitivity, robustness, and solid-state nature. To ensure linear response and accurate angular velocity measurement, a fixed $\pi/2$ phase bias is typically introduced between the clockwise (CW) and counter-clockwise (CCW) beams using active modulation components. However, these active elements increase system complexity, power consumption, cost, and susceptibility to thermal drift and long-term degradation. In this work, we present a novel IFOG configuration that, to the best of our knowledge, achieves—for the first time—simultaneous operation at two quadrature points ($\pi/2$ and $3\pi/2$), providing natural noise suppression without relying on active components. This is made possible through the integration of a Non-Reciprocal Polarization-Dependent Phase Shifter (NRPPS), which introduces a pure $\pi/2$ passive phase shift. We detail the optical architecture and provide theoretical modeling using Jones calculus to demonstrate how the NRPPS element introduces passive quadrature biasing. Simulation results show that the proposed NRPPS-IFOG achieves significantly improved sensitivity, with Angular Random Walk (ARW) values up to 40× lower than those of conventional IFOGs, depending on the fiber coil length. The design further leverages quadrature-phase signal detection and adjustable temporal offsets between photodetectors to enhance noise suppression. This passive biasing approach eliminates the need for active modulation, offering reduced power consumption, improved stability, and enhanced long-term reliability.
\end{abstract}

\keywords{IFOG \and gyroscope \and passive \and bias \and non-reciprocity}

\section{Introduction}

Interferometric Fiber Optic Gyroscopes (IFOGs) have become a cornerstone in modern precision navigation systems due to their exceptional sensitivity \cite{sanders1996fiber}, reliability \cite{lefevre2013fiber}, and ability to function without moving parts \cite{grattan2000fiber}. These gyroscopes rely on the Sagnac effect \cite{sagnac1913ether}, where light is split into two beams traveling in opposite directions through a fiber optic coil as clockwise (CW) and counter-clockwise (CCW). Rotation causes a phase shift between the two counter-propagating beams, and this phase difference is directly proportional to the angular velocity of the system. The phase shift is detected through interference at a photodetector, providing a highly accurate measurement of rotation. To ensure precise measurement of both the magnitude and direction of rotation, a fixed phase bias—typically at $\pi/2$—is introduced between the two beams. This phase shift moves the system’s response from its maximum or minimum point, placing it in the linear region of the output response curve, where sensitivity is maximized. Traditionally, active modulation components, such as electro-optic and piezoelectric phase modulators, are used to achieve this bias. These components modulate the optical path by applying a controlled voltage, shifting the phase at critical points in the light cycle \cite{lefevre2022fiber}.

However, the reliance on active components introduces significant challenges. Active modulators demand complex control electronics, dedicated power supplies, and thermal stabilization \cite{kiraci2017temperature, wang2014compensation, mao2005research} to ensure reliable performance under varying environmental conditions. These requirements not only increase the overall size and power consumption of the system but also add to its cost. Moreover, active components are prone to reliability issues such as aging, drift on phase \cite{bi2018influence,wang2010study, sun2010study}, and mechanical failure, which can compromise long-term performance and require frequent recalibration or replacement. Additionally, for longer fiber version of fiber optic gyroscopes which are needed for higher sensitivity, although, there are some works on surpassing thermal phase noise (TPN) \cite{takei2023simultaneous}, TPN is a still key parameter for the sensitivity limitations \cite{song2017modeling}. Additionally, devices like piezoelectric phase shifters are limited by practical frequency constraints especially short fiber coil, which in turn limit the size and design flexibility of the system \cite{trommer1990passive, trommer1996passive}.

While these issues are evident in IFOGs, they are not confined to this specific gyroscope type. They also apply to a broader range of optical gyroscope systems, where active phase biasing is crucial to system operation. Therefore, there is a pressing need for a passive, robust, and low-maintenance phase biasing solution that can overcome the limitations of active modulation in IFOGs and also other optical gyroscope systems.

Several alternative approaches have been explored to address these challenges, with some success in utilizing passive biasing techniques. However, these solutions often come with trade-offs in terms of complexity, sensitivity, and reliability. Early work, such as that by Sheem in 1980, explored the use of a 3×3 directional coupler to replace traditional beam splitters, eliminating the need for active modulation \cite{sheem1980fiber}. While this approach avoided some of the issues associated with active components, it introduced new challenges, including sensitivity to asymmetry between output ports, increased signal loss, and greater complexity in signal processing.

Later advancements, including the work by Kajioka and Matsumura in 1984, sought to reduce the need for active modulation by detecting changes in polarization states between two light beams traveling through the fiber coil \cite{kajioka1984single}. Rotation causes a change in the polarization state, which results in an interference signal at the detector, with the difference in detected power proportional to the rotation rate. While such systems showed promise, they remained susceptible to environmental fluctuations and component variability, which could compromise accuracy \cite{doerr1994orthogonal}. Other designs, such as those introduced by Huang in 2007 and Jabo and Xiao Qian in 2009, aimed to achieve passive biasing through novel optical fiber couplers \cite{huang2010passively} or branching units \cite{Bo2009}, respectively, but these methods often came with their own set of challenges, such as complex fabrication and precise tuning requirements.

Most recently, Hakimi et al. in 2024 proposed a system utilizing a topological phase bias element to introduce a passive phase shift between the counter-propagating beams \cite{hakimi2024passive}. While this system showed improvements in reducing system complexity, it faced issues related to polarization mode dispersion and drift, especially when non-polarization-maintaining fibers were used. These drawbacks suggest that fully realizing a passive solution for high-precision, robust optical gyroscopes remains an ongoing challenge.

None of the mentioned passive techniques has fully addressed the need for a high-performance, low-maintenance, and reliable phase-biasing solution for optical gyroscopes to work around $\pi/2$. This paper explores the potential for a new passive modulation scheme that aims to overcome the limitations of current systems, providing a more robust and versatile solution for IFOGs and other interferometric gyroscopes.

In 2014, Steve Yao’s energy-efficient optic gyroscope indeed relies on optical interference principles to detect rotation around $\pi/2$ passively. The design uses a polarization beam splitter (PBS) and a retarder to split and recombine light beams traveling in opposite directions within a fiber optic loop \cite{yao2016energy}. The retarder introduces a controlled phase bias (around $\pi/2$) between the two beams, positioning the system at its point of maximum sensitivity to rotation-induced phase differences. This configuration converts the phase shift caused by the Sagnac effect into changes in interference intensity. 

In this paper, while Yao’s work demonstrated the first IFOG system operating around a $\pi/2$ phase bias, we introduce and investigate a method for achieving $\pi/2$ passive biasing in Interferometric Fiber Optic Gyroscopes (IFOGs) that, for the first time, enables simultaneous operation at two quadrature points ($\pi/2$ and $3\pi/2$) and provides spontaneous noise suppression. In the first section, we present a concise comparison between a conventional IFOG setup and the proposed NRPPS-IFOG system, highlighting the key differences in system configuration. The second section delves into the working mechanism of the Non-Reciprocal Polarization Dependent Phase Shifter (NRPPS) element, explaining how it introduces passive phase biasing between the clockwise (CW) and counter-clockwise (CCW) light beams within the IFOG. Finally, we examine the performance of the NRPPS-IFOG system through simulations, comparing it with a conventional IFOG in terms of sensitivity, accuracy, and long-term stability. The simulation results demonstrate the effectiveness of the proposed approach, highlighting its potential for enhanced stability, sensitivity, and noise suppression in practical applications.

\begin{figure}[!t]
\centering
\includegraphics[width=0.95\textwidth]{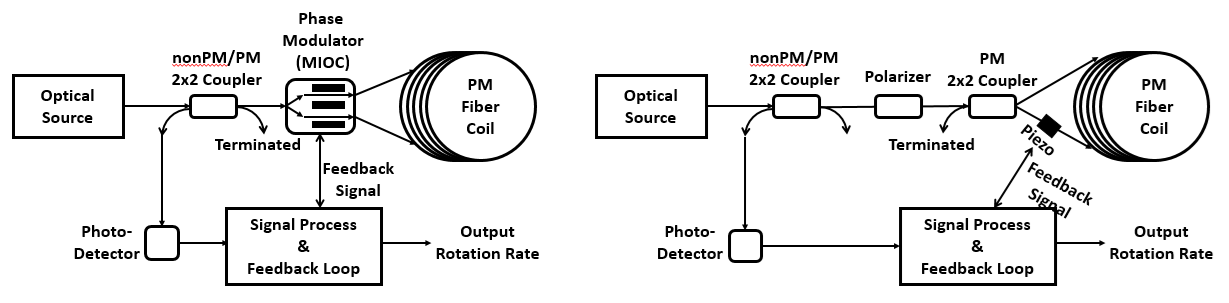}
\caption{\label{fig:IFOG_SetupComparison} (Left) A multi-integrated optical circuit (MIOC) integrated conventional IFOG, (right) Piezoelectric ceramic actuator integrated conventional IFOG.}
\end{figure}

In its minimum configuration, an active-biased interferometric fiber optic gyroscope (IFOG) consists of a coherent light source, a beam splitting and recombination unit, a phase modulation element, a sensing coil, and a photodetector system, all connected through polarization-maintaining (PM) fiber as seen in Fig-~\ref{fig:IFOG_SetupComparison}.

Specifically, light from the coherent source is fed into a fiber optic coupler or splitter, which directs the beams into two counter-propagating paths around the PM fiber sensing coil. After traversing the coil, the beams return and recombine at the coupler, where the resulting interference signal is directed to the photodetector.
To linearize the system’s response near zero rotation, a $\pi/2$ phase bias is introduced between the clockwise (CW) and counter-clockwise (CCW) beams. This is typically achieved using one of two common active modulation techniques:

        \begin{enumerate}
            \item Electro-optic modulation, implemented via a MIOC that integrates polarizer, phase modulator, and Y-junction functions, with a feedback loop controlling the phase modulation signal Fig-~\ref{fig:IFOG_SetupComparison}-left.
            \item Piezoelectric modulation, where a section of the optical fiber is wrapped around or bonded to a piezoelectric ceramic actuator that induces a controllable phase shift through mechanical deformation Fig-~\ref{fig:IFOG_SetupComparison}-right.
        \end{enumerate}
        
In both cases, the system relies on active feedback electronics to maintain the bias and process the detected signal, producing an angular velocity output.

\begin{figure}[!b]
\centering
\includegraphics[width=0.8\textwidth]{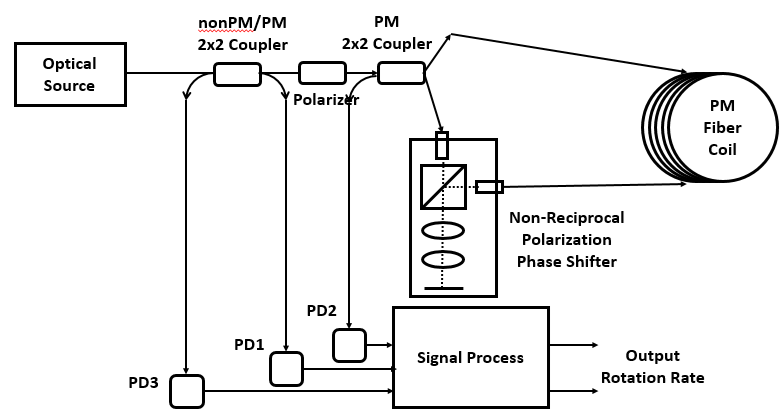}
\caption{\label{fig:NRPPS-IFOG_Setup} A Non-Reciprocal Polarization Phase Shifter based IFOG system.}
\end{figure}

As seen in Fig-~\ref{fig:NRPPS-IFOG_Setup}, a NRPPS-based IFOG system uses a passive, non-reciprocal polarization phase-shifting mechanism to replace conventional active modulators. Actually, NRPPS is not a new concept and by using this kind of element, stable and robust 9-figure PM Yb:fiber mode-locked oscillator based on a nonlinear amplifying loop mirror (NALM) can be achieved \cite{jiang2016all}. The NRPPS-IFOG setup starts with a coherent light source connected to an optional nonPM or PM 2×2 coupler, depending on whether the source is linearly polarized. A polarizer ensures the beam is linearly polarized before entering the main PM 2×2 coupler, which splits the light into clockwise (CW) and counterclockwise (CCW) paths.
The CW beam, initially horizontally polarized, passes through a PM collimator, a polarization beam splitter (PBS), a passive non-reciprocal element in terms of polarization that rotates its polarization by +45°, a retarder aligned to maintain its phase, and then reflects back through the same path. After completing the round trip, its polarization rotates by a net +90° (vertical), allowing proper recombination at the PBS and couples to the second PM collimator which allows the beam travels into the PM sensing coil and re-entry into the PM 2x2 coupler.

The CCW beam follows a symmetric but mirror-inverted path, passing through the sensing coil, collimator, PBS, non-reciprocal element, retarder (introducing a phase shift $\phi_{r}$), mirror, and back, accumulating an additional +90° polarization rotation and a total phase shift of $\pi/2 + 2\phi_{r}$ relative to its starting point. At the final coupler stage, two outputs are measured (for feedback, PD1 can also be used):
PD2 is placed at the output of PM 2x2 coupler and records interference between CW ($0\pi$) and CCW ($\pi + 2\phi_{r}$), yielding a relative phase shift of $\pi + 2\phi_{r}$. The sum of the CW and CCW amplitudes with shifted by Sagnac phase $\phi_{i}$ for PD2;

\begin{eqnarray}
\centering
&&\label{eqnarray:1} E_{PD2} = E_{CW}e^{0j} + E_{CW}e^{-j(\phi_i+(\pi+2\phi_{r}))},\\
assume
&&\label{eqnarray:2} E_{CW}=E_{CCW}=E,\\
&&\label{eqnarray:3}E_{PD2}=E(\cos{(0^\circ)}+j\sin{(0^\circ)}+\cos{(\phi_{i}+(\pi + 2\phi_{r}))}-j\sin{(\phi_{i}+(\pi + 2\phi_{r}))}).
\end{eqnarray}

The intensity on the PD2 detector becomes;

\begin{eqnarray}
\centering
&&\label{eqnarray:4} I_{PD2} = |E_{PD2}|^2 = 2I(1+\cos{(\phi_{i}+(\pi + 2\phi_{r}))}).
\end{eqnarray}

PD3 which is placed after the (optional) nonPM/PM coupler and captures the interference between CW ($\pi/2$) and CCW ($\pi/2 + 2\phi_{r}$), yielding a relative phase shift of $2\phi_{r}$. The sum of the CW and CCW amplitudes with shifted by Sagnac phase $\phi_{i}$ for PD3;

\begin{eqnarray}
\centering
&&\label{eqnarray:5} E_{PD3} = E_{CW}e^{0j} + E_{CW}e^{-j(\phi_i+2\phi_{r})},\\
assume
&&\label{eqnarray:6} E_{CW}=E_{CCW}=E,\\
&&\label{eqnarray:7} E_{PD3}=E(\cos{(0^\circ)}+j\sin{(0^\circ)}+\cos{(\phi_{i}+ 2\phi_{r})}-j\sin{(\phi_{i}+2\phi_{r}}).
\end{eqnarray}

The intensity on the PD3 detector becomes;

\begin{eqnarray}
\centering
&&\label{eqnarray:8} I_{PD3} = |E_{PD3}|^2 = 2I(1+\cos{(\phi_{i}+ 2\phi_{r})}).
\end{eqnarray}

\begin{figure}[!b]
\centering
\includegraphics[width=0.8\textwidth]{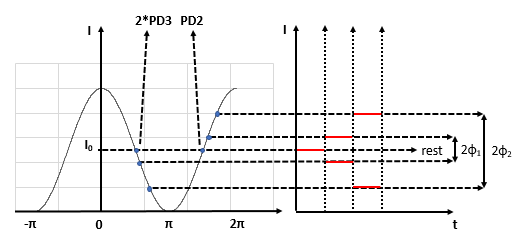}
\caption{\label{fig:Sinus_Response} Output signals of PD2 and PD3 section for NRPPS-IFOG while rest and with rotation $\phi_{1}$ and $\phi_{2}$.}
\end{figure}

If the retarder is set as an 8-wave plate at 45° which means $\phi_{r}=\pi/4$,  the system outputs signals in quadrature ($\pi/2$ and $3\pi/2$ as seen in Fig-~\ref{fig:Sinus_Response}), ensuring linear sinusoidal response, complementary detection, and enhanced sensitivity without active modulation. According to user requirements, the retarder can be adjusted to other values, but setting it at $\phi_{r}=\pi/4$ ensures a phase shift of $\pi/2$. The intensity of the PD2 and PD3 is different due to different locations of the system and without considering the losses of the components, the PD3 will be half of the PD2 level as seen in Fig-~\ref{fig:Sinus_Response}.

From this point onward in the paper, the NRPPS scheme will be analyzed using a configuration where the NRPPS incorporates an 8-wave plate retarder oriented to match between the polarization of CW beam and the fast axis of the retarder. First, the Jones matrix definitions will be given for the optics in terms of polarization and phase then for CW and CCW beams are calculated over these definitions according to NRPPS design.

\begin{eqnarray}
\centering
\textbf{horizontal polarization}  
&&\label{eqnarray:9} 
  H =
  \left[ {\begin{array}{cc}
    1 \\
    0 \\
  \end{array} } \right],\\
  \textbf{vertical polarization}
&&\label{eqnarray:10} 
  V =
  \left[ {\begin{array}{cc}
    0 \\
    1 \\
  \end{array} } \right],\\
  \textbf{PBS}
&&\label{eqnarray:11} 
  J_{PBS} =
  \left[ {\begin{array}{cc}
    1 & 0\\
    0 & 1\\
  \end{array} } \right],\\
  \textbf{Rotation matrix}
&&\label{eqnarray:12} 
  R(\theta) =
  \left[ {\begin{array}{cc}
    cos(\theta) & -sin(\theta)\\
    sin(\theta) & cos(\theta)\\
  \end{array} } \right],\\
    \textbf{Collimator rotation}
&&\label{eqnarray:13} 
  J_{coll}(\theta) =
  \left[ {\begin{array}{cc}
    cos(\theta) & -sin(\theta)\\
    sin(\theta) & cos(\theta)\\
  \end{array} } \right],\\
 \textbf{Passive Non-Reciprocal Element at $R(45^o )$}
&&\label{eqnarray:14} 
  J_{non-reciprocal El.} = 1/\sqrt{2}
  \left[ {\begin{array}{cc}
    1 & -1\\
    1 & 1\\
  \end{array} } \right],\\
\textbf{Retarder at $R(R_1)$}
&&\label{eqnarray:15} 
  J_{retarder} = R(-R_1 )
  \left[ {\begin{array}{cc}
    e^{i\phi_r} & 0\\
    0 & e^{-i\phi_r}\\
  \end{array} } \right]R(R_1 ),\\
    \textbf{mirror}
&&\label{eqnarray:16} 
  J_{mirror} =
  \left[ {\begin{array}{cc}
    1 & 0\\
    0 & -1\\
  \end{array} } \right].
\end{eqnarray}

\begin{figure}[!b]
\centering
\includegraphics[width=0.8\textwidth]{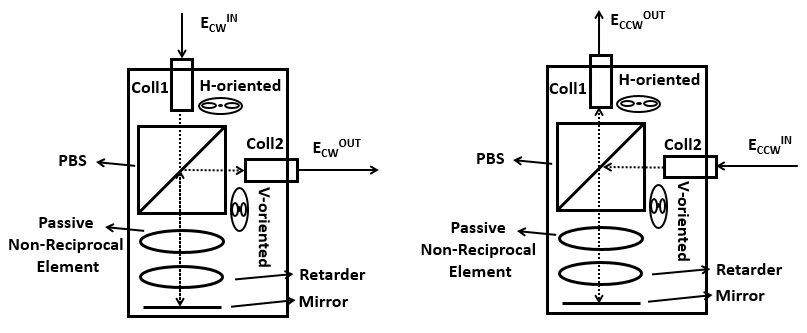}
\caption{\label{fig:CW_CCW_NRPPS} (left) Non-Reciprocal Polarization Phase Shifted Design with respect to CW beam, (right) Non-Reciprocal Polarization Phase Shifted Design with respect to CCW beam.}
\end{figure}

Fig-~\ref{fig:CW_CCW_NRPPS}-(left) illustrates how the Non-Reciprocal Polarization Phase Shifted Design is working with respect to CW beam. For the CW beam which comes from one port of the 2x2 PM coupler, the light is collimated by using a PM fiber collimator and aligned horizontally relative to the polarizing beam splitter (PBS). It passes directly through the PBS toward the passive non-reciprocal element, where it undergoes a +45° polarization rotation. Reaching the 8-wave plate retarder, which is oriented to match between the polarization of CW beam and the fast axis of the retarder, no additional relative phase shift is introduced. The beam then reflects off the mirror and retraces its path, again passing through the retarder without gaining phase shift. As it passes back through the non-reciprocal element, it experiences another +45° polarization rotation, totaling +90° and effectively changing its polarization to vertical. This vertical polarization causes the beam to be rerouted by the PBS into a different output path, where it couples into another polarization-maintaining (PM) fiber collimator aligned 90° relative to the PBS, realigning the beam horizontally and guiding it forward as the CW output.

Let’s apply the Jones matrix according to the CW beam path in NRPPS design:

\begin{eqnarray}
\centering
&&\label{eqnarray:17} 
  E_{in}^{CW}=H =
  \left[ {\begin{array}{cc}
    1 \\
    0 \\
  \end{array} } \right],\\
  &&\label{eqnarray:18} 
 J_{coll2}(90^\circ)*J_{PBS}*J_{non-reciprocal El.}(45^\circ)*J_{retarder}(R_1)*J_{mirror}*J_{retarder}(R_1)*  \nonumber\\
 && *J_{non-reciprocal El.}(45^\circ)* J_{PBS}*J_{coll1}(0^\circ)*E_{in}^{CW} =E_{out}^{CW},\\
 &&\label{eqnarray:19} 
  E_{out}^{CW}=H =
  \left[ {\begin{array}{cc}
    1 \\
    0 \\
  \end{array} } \right].
\end{eqnarray}

Fig-~\ref{fig:CW_CCW_NRPPS}-(right) illustrates how the NRPPS design is working with respect to CCW beam. For the CCW beam which comes from PM sensing coil, the light is collimated by using a PM fiber collimator and aligned vertically relative to the PBS. It is reflected by the PBS toward the passive non-reciprocal element, where it undergoes a +45° polarization rotation. When it reaches the 8-wave plate retarder, which is oriented +45° relative to horizontal such that the slow axis of the retarder matches with the polarization of CCW beam and the beam experiences a $\pi/4$ phase shift relative to the CW beam. After reflecting off the mirror and retracing its path, the beam passes through the retarder again, adding another phase shift of the same magnitude, resulting in a total phase shift of $\pi/2$. As it passes back through the non-reciprocal element, it gains an additional +45° polarization rotation, reaching a net +90° and returning to horizontal polarization. This polarization change causes the PBS to route the beam through the alternate output path, where it couples into a PM fiber collimator aligned 0° relative to the PBS, ensuring the beam exits horizontally aligned as the CCW output.

\begin{figure}[!b]
\centering
\includegraphics[width=0.45\textwidth]{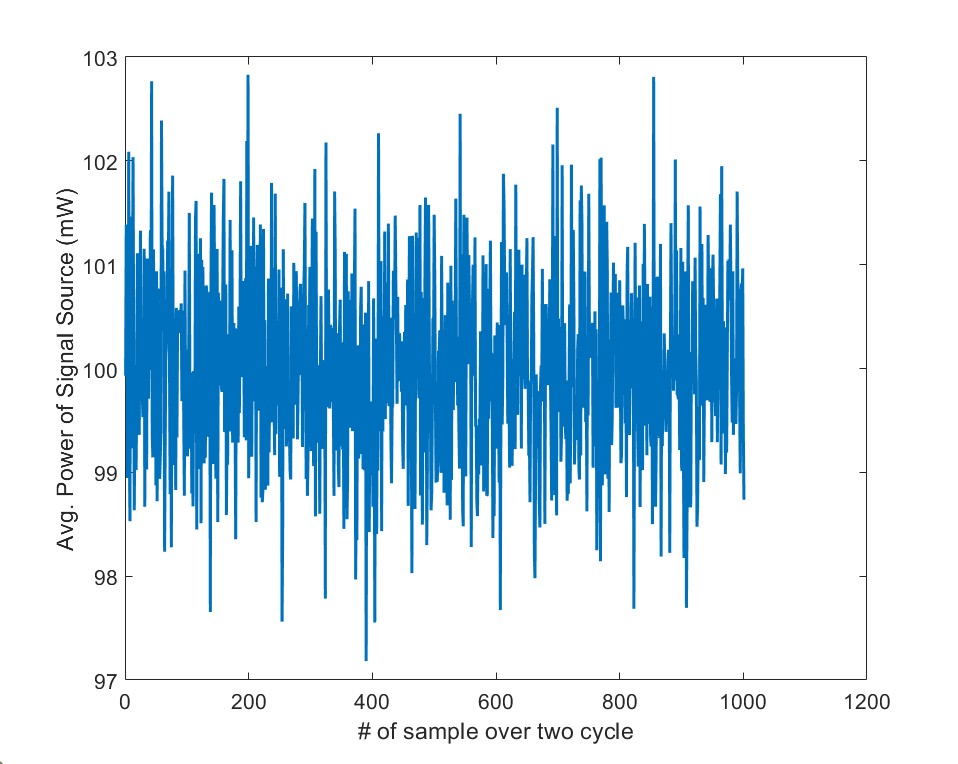}
\includegraphics[width=0.45\textwidth]{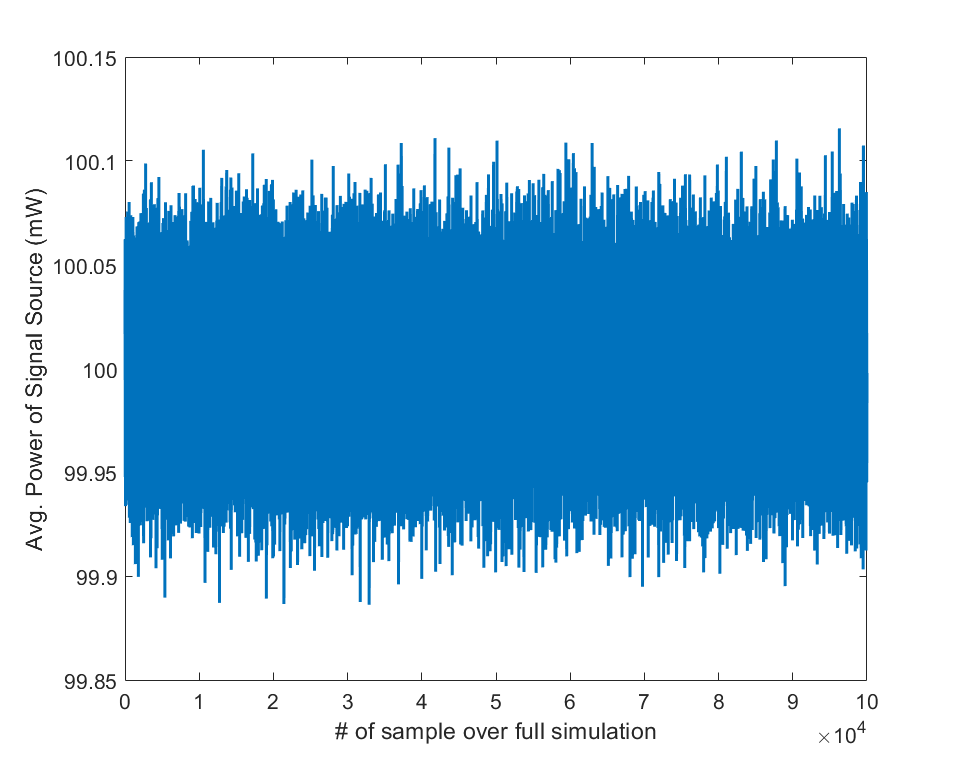}
\caption{\label{fig:PowerIn} (left) The input signal source for conventional-IFOG simulation as number of samples in 2 cycles, (right) number of samples over full simulation (right) for 500m. long fiber coil.}
\end{figure}

Let’s apply the Jones matrix according to the CCW beam path in NRPPS design:

\begin{eqnarray}
\centering
&&\label{eqnarray:20} 
  E_{in}^{CCW}=H =
  \left[ {\begin{array}{cc}
    1 \\
    0 \\
  \end{array} } \right],\\
  &&\label{eqnarray:21} 
 J_{coll1}(0^\circ)*J_{PBS}*J_{non-reciprocal El.}(45^\circ)*J_{retarder}(R_1)*J_{mirror}*J_{retarder}(R_1)*  \nonumber\\
 && *J_{non-reciprocal El.}(45^\circ)* J_{PBS}*J_{coll2}(90^\circ)*E_{in}^{CCW} =E_{out}^{CCW},\\
 &&\label{eqnarray:22} 
  E_{out}^{CW}=e^{j2\phi_r}H = e^{j2\phi_r}
  \left[ {\begin{array}{cc}
    1 \\
    0 \\
  \end{array} } \right].
\end{eqnarray}

For example, if $\phi_r= \pi/4$ which represents 8-wave retarder, the total phase shift:

\begin{eqnarray}
\centering
&&\label{eqnarray:23} 
  E_{out}^{CW}=
  \left[ {\begin{array}{cc}
    1 \\
    0 \\
  \end{array} } \right],\\
 &&\label{eqnarray:24} 
  E_{out}^{CW}=e^{j2\phi_r}
  \left[ {\begin{array}{cc}
    1 \\
    0 \\
  \end{array} } \right]=e^{j\pi/2}
  \left[ {\begin{array}{cc}
    1 \\
    0 \\
  \end{array} } \right].
\end{eqnarray}

As a result, the NRPPS introduces a precise $\pi/2$ phase shift between $E_{out}^{CW}$  and $E_{out}^{CCW}$, effectively shifting the operating point into the linear region of the sinusoidal response. 

The simulations were carried out using a custom-developed MATLAB code. The optical source employed had a center wavelength of 1550 nm, a 20 nm Gaussian-shaped bandwidth, an average power of 100 mW as seen in Fig.-~\ref{fig:PowerIn}. 

For conventional IFOG, a multifunction integrated optical circuit (MIOC) served as a 2×2 polarization-maintaining (PM) coupler, a polarizer, and a modulator is used. Modulation was applied without introducing additional noise. To suppress transient effects associated with rising and falling edges, $98\%$ of the signal was utilized for integration. In the conventional IFOG configuration, two modulation cycles are required to calculate the rotation rate, resulting in one data point being obtained every two cycles after averaging the whole both cycles points.

For the NRPPS-IFOG, the same optical source was employed. A polarizer with a polarization extinction ratio (PER) of 30 dB was placed before the 2×2 PM coupler connected to the sensing coil. An 8-wave retarder was incorporated in the NRPPS section to achieve a $\pi/2$ phase shift at the output. As illustrated in the system diagram, there are three output ports as two for extracting rotation-induced phase shifts and one for monitoring the input beam. Photodetectors in Fig.-~\ref{fig:NRPPS-IFOG_Setup}, PD2 and PD3 were used to calculate the rotation rate, with a 3-meter fiber length difference introduced to create a temporal offset between the signals. Otherwise, perfect temporal match makes the system noiseless, which will be addressed for the future papers in terms of detailed investigation of the NRPPS-IFOG system. In the NRPPS-IFOG configuration, a key advantage is the continuous rotation measurement, allowing one data point to be acquired per modulation cycle after averaging the whole cycle points. 

\begin{table}
\centering
\caption{(left) Performance comparison of conventional IFOG and NRPPS-IFOG in terms of rotation rate and (right) Overlapping Allan variance analysis for 200m., 1000m. and, 2000m. fiber spool.}
\label{tab:table1}
\begin{tabular}{|c|c|c|}
\hline
\parbox{3cm}{$200m.$ \\
\\
ARW (conv.):  \\
$0.0083 ^\circ/\sqrt{hr}$ \\
\\
ARW (NRPPS): \\
$0.0007 ^\circ/\sqrt{hr}$ \\} 
  & \raisebox{-0.5\height}{\includegraphics[height=4cm]{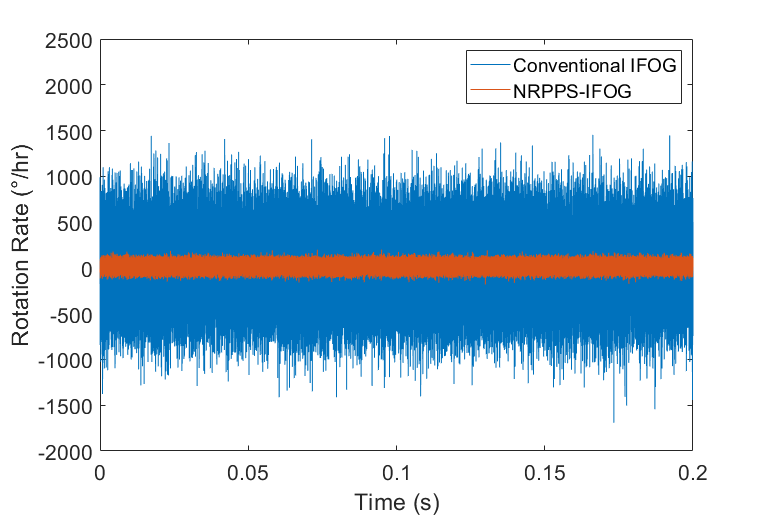}} 
  & \raisebox{-0.5\height}{\includegraphics[height=4cm]{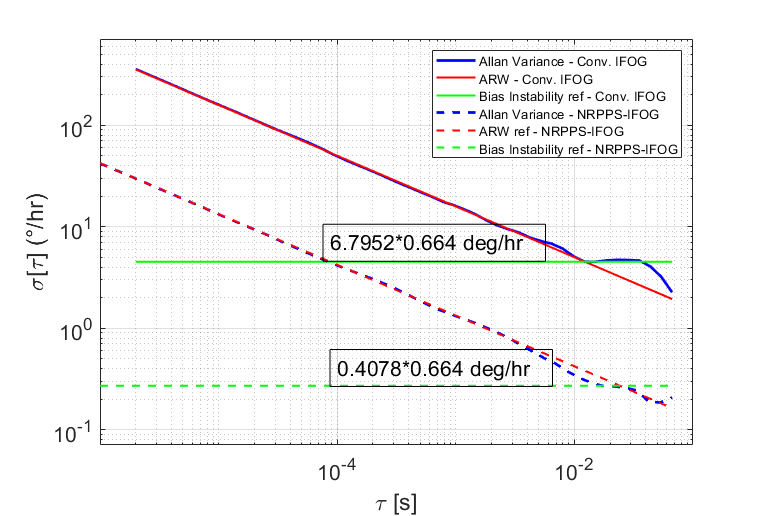}} \\
  \parbox {3cm}{$1000m.$ \\
  \\
ARW (conv.):  \\
$0.0017 ^\circ/\sqrt{hr}$ \\
\\
ARW (NRPPS): \\
$0.00006 ^\circ/\sqrt{hr}$ \\} 
  & \raisebox{-0.5\height}{\includegraphics[height=4cm]{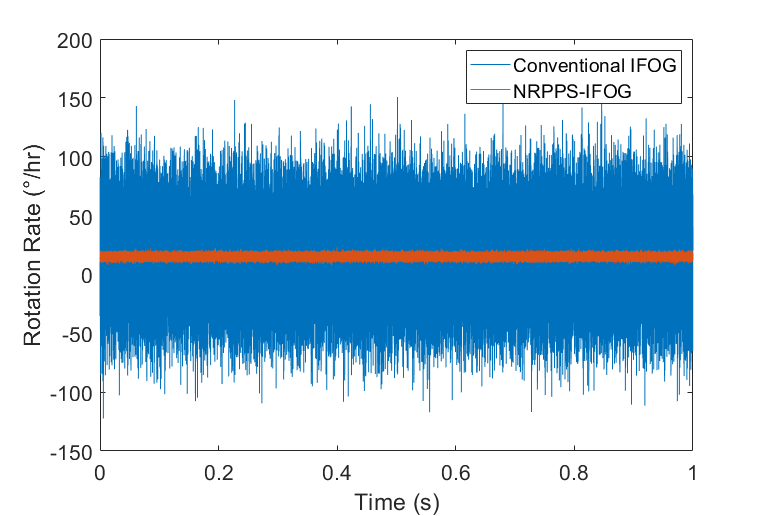}} 
  & \raisebox{-0.5\height}{\includegraphics[height=4cm]{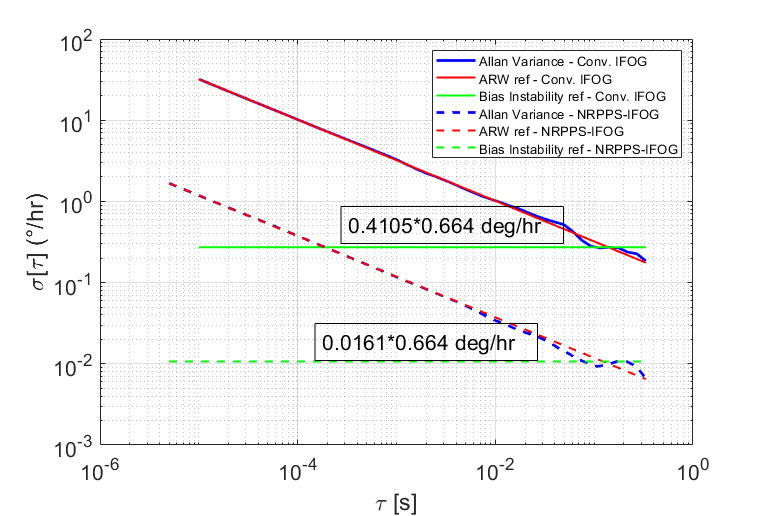}} \\
    \parbox {3cm}{$2000m.$ \\
    \\
ARW (conv.):  \\
$0.0008 ^\circ/\sqrt{hr}$ \\
\\
ARW (NRPPS): \\
$0.00002 ^\circ/\sqrt{hr}$ \\} 
  & \raisebox{-0.5\height}{\includegraphics[height=4cm]{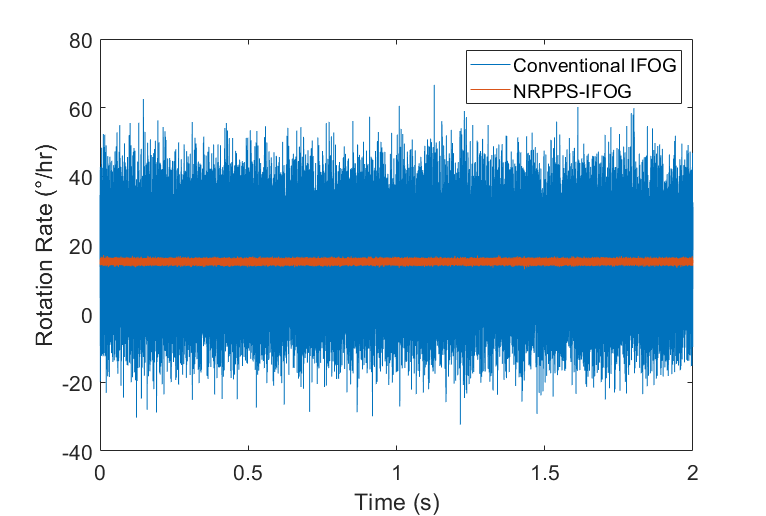}} 
  & \raisebox{-0.5\height}{\includegraphics[height=4cm]{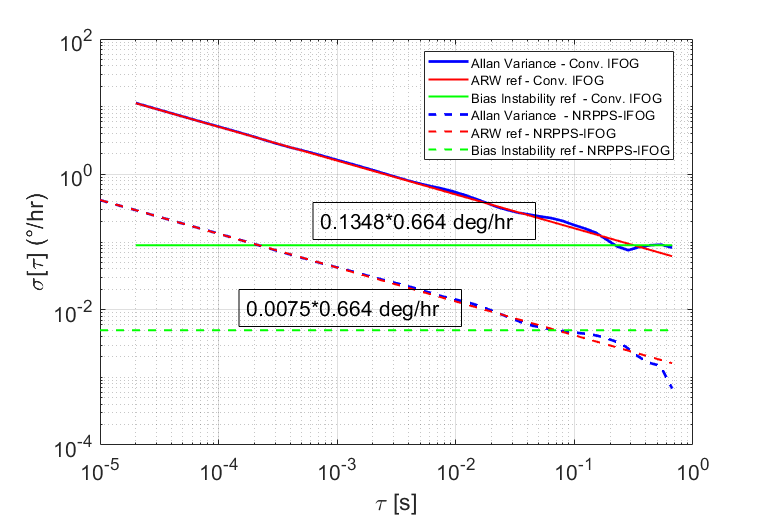}} \\
\hline
\end{tabular}
\end{table}

In the simulation, sensitivity was calculated based on 200000 round trips through the sensing coil, assuming a constant rotation rate equal to Earth's rotation (15°/s). During each coil cycle, one data 
sample is read every ~1/c – where c is speed of light, and this sampling interval remains consistent across all system configurations. For conventional IFOG, one data point (the avg. of the output) is collected over two coil cycles, whereas the NRPPS-IFOG configuration provides one data point due to its ability to extract one measurement per coil cycle. To evaluate the effect of polarization-maintaining (PM) coil length on system performance, simulations were conducted using coil lengths of $200m.$, $1000m.$, and $2000m.$. with all fiber spools having a diameter of 10cm., as seen Table-~\ref{tab:table1}.

\begin{table}
\centering
\caption{Performance comparison of the NRPPS-IFOG in terms of temporal offset between the signals of PD3 and PD2 for 2000m. fiber coil.}
\label{tab:table2}
\begin{tabular}{|c|c|c|}
\hline
\parbox{3cm}{Rotation Rates} 
  & \raisebox{-0.5\height}{\includegraphics[height=6cm]{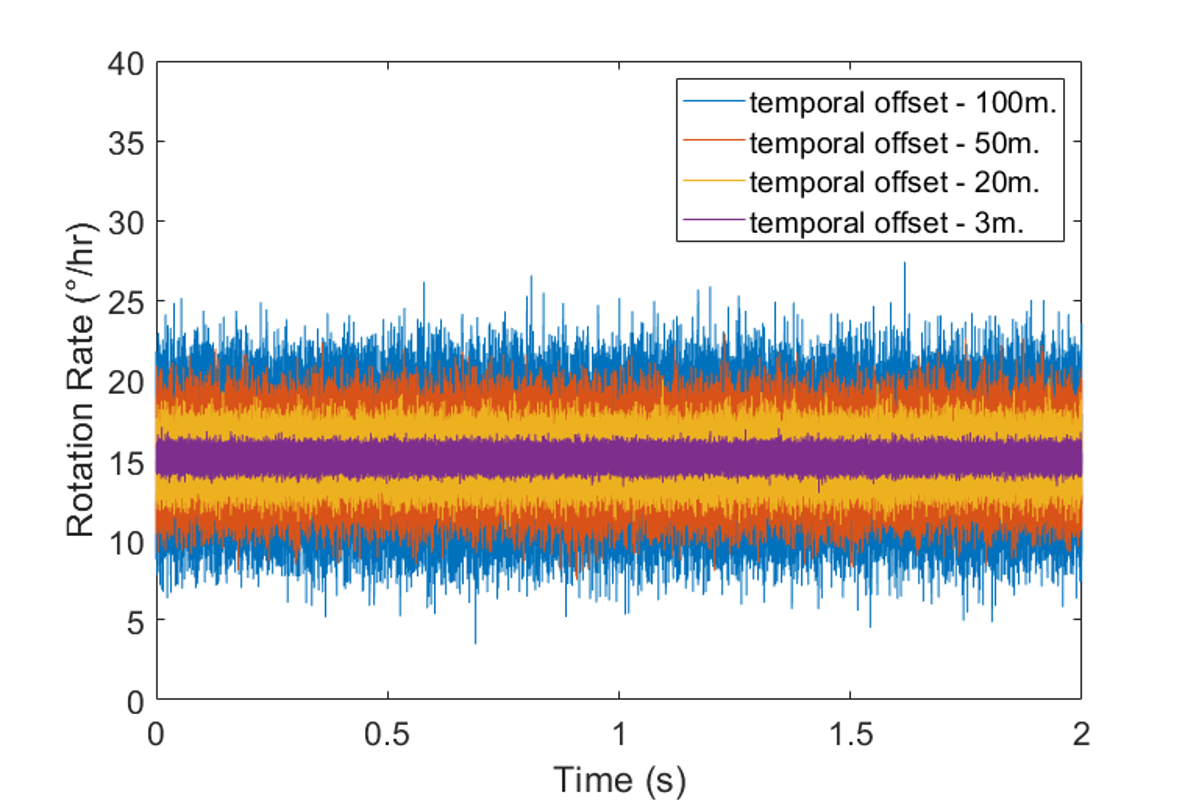}} \\
  \parbox {3cm}{$ARW - 3m.$ \\
$0.000022 ^\circ/\sqrt{hr}$ \\
\\
$ARW - 20m.$ \\
$0.000056 ^\circ/\sqrt{hr}$ \\
\\
$ARW - 50m.$ \\
$0.000090 ^\circ/\sqrt{hr}$ \\
\\
$ARW - 100m.$ \\
$0.000133 ^\circ/\sqrt{hr}$ \\} 
  & \raisebox{-0.5\height}{\includegraphics[height=6cm]{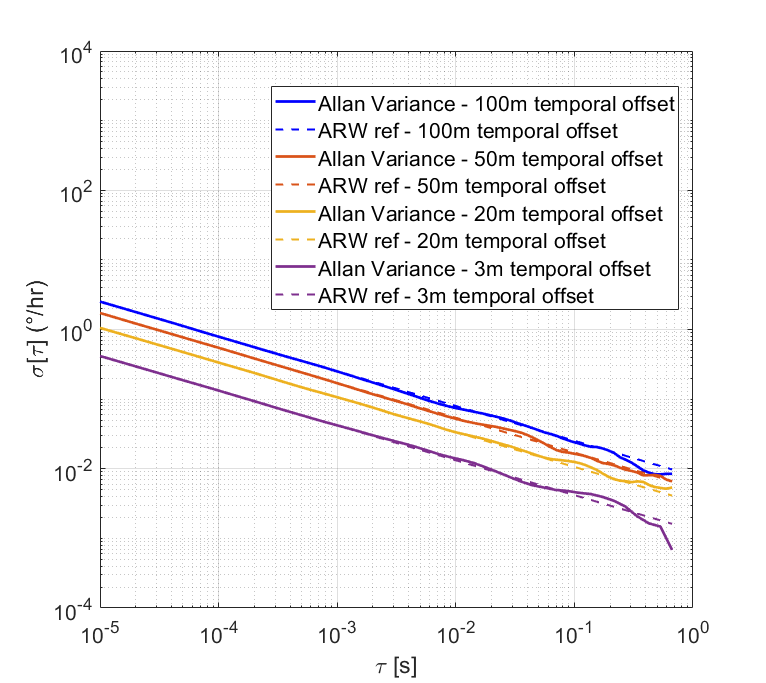}} \\
\hline
\end{tabular}
\end{table}

As a result, we observed that the Angular Random Walk (ARW) values for the NRPPS-IFOG are approximately 10×, 30×, and 40× lower than those of the conventional IFOG for fiber lengths of $200m.$, $1000m.$, and $2000m.$, respectively. This significant improvement in sensitivity is primarily attributed to the use of PD2 and PD3, which are actually same signal but receive $\pi$-phase-shifted quadrature components of the sinusoidal signal. Additionally, noise cancellation can be enhanced by adjusting the temporal offset between the signals from PD2 and PD3. Our simulations indicate 
that the performance of the NRPPS-IFOG is strongly influenced by this temporal offset, as summarized in Table-~\ref{tab:table2}. Since the primary focus of this paper is to introduce and explain the operating principle of the NRPPS-IFOG, a detailed analysis of the temporal offset’s impact on performance will be presented in future work.
\section{Conclusion}
In conclusion, the NRPPS-IFOG, provides a purely passive $\pi/2$ phase modulation for the first time based on our knowledge- has been analytically studied in comparison with a conventional IFOG system. The performance of the NRPPS-IFOG demonstrates a significant and favorable improvement with Angular Random Walk (ARW) values up to 40× lower than those of
conventional IFOGs, depending on the fiber coil length. Although this study compares the NRPPS-IFOG only with the conventional IFOG, the proposed element has potential applications in other optical gyroscopes that require active biasing. Thus, it opens a new avenue for optical gyroscope systems in terms of;

        \begin{enumerate}
            \item To provide a flexible, implementation-independent concept — adaptable to various passive, non-reciprocal mechanisms (e.g., magneto-optic, photonic, or topological elements).
            \item To provide a passive optical phase biasing system — eliminates the need for active modulators, such as electro-optic phase shifters, not only for IFOG but in all gyroscope systems requiring a controlled phase shift between counter-propagating or interacting optical beams.
            \item To enable full optical functionality without active electronic control — simplifying the system and reducing reliance on high-speed electronics.
            \item 	To reduce power consumption — by avoiding active components that generate or respond to electrical signals.
            \item To enhance long-term system reliability and stability — especially in environments where active modulation may degrade performance over time.
            \item To support implementation in environmentally demanding platforms — such as aerospace, marine, and autonomous navigation systems where passive solutions are advantageous.
            \item To support continuous gyroscope measurement — by enabling uninterrupted readout without duty-cycle constraints imposed by modulated biasing schemes.
            \item To allow for unrestricted PM fiber coil length — permitting flexible design and scaling of the sensing coil without phase bias limitations imposed by modulation timing.  
            \item To enable chip-scale integration of the passive biasing element — allowing the development of ultra-compact gyroscope systems that occupy minimal physical space, supporting miniaturization for emerging photonic and integrated sensor platforms.  
        \end{enumerate}

Additionally, by placing a piezoelectric ceramic actuator on the backside of the mirror, the system can be modulated to implement a feedback mechanism to make the system closed loop. However, since this paper focuses solely on the passive-biased configuration of the NRPPS-IFOG, the modulated version will be investigated in future work. Furthermore, while the NRPPS-IFOG demonstrates enhanced performance, it is well known that optical components used in the NRPPS element—such as the non-reciprocal element and the phase retarder—are sensitive to temperature variations. This thermal dependence which can be stabilized by temperature controller, can influence overall system performance and will also be addressed in future studies.

\section*{Acknowledgments}
A provisional patent application has been filed for the technology described in this paper. O.A. conducted the simulations, analyzed the results and supervised the manuscript. M.G.A. reviewed the manuscript.

\bibliographystyle{unsrt}
\bibliography{NRPPS-IFOG}

\end{document}